# Taming active turbulence with patterned soft interfaces

P. Guillamat, J. Ignés-Mullol, and F. Sagués*

Department of Materials Science and Physical Chemistry and Institute of Nanoscience and Nanotechnology (IN2UB), University of Barcelona, Martí i Franquès 1, 08028 Barcelona. Catalonia, Spain.

27-Oct-16

**Abstract**

Active matter embraces systems that self-organize at different length and time scales, often exhibiting turbulent flows. Here, we use a quasi-two-dimensional nematically ordered layer of a protein-based active gel to experimentally demonstrate that the geometry of active flows, which we have characterized by means of the statistical distribution of eddy sizes, can be reversibly modified. To this purpose, the active material is prepared in contact with a thermotropic liquid crystal, whose lamellar smectic-A phase tiles the water/oil interface with a lattice of anisotropic domains that feature circular easy-flow directions. The active flow, which arises from the motion of parabolic folds within the filamentous active material, becomes effectively confined into circulating eddies that are in registry with the underlying smectic-A domains. Based on topological arguments applied to the circular confinement, together with well-known properties of the active material, we show that the structure of the active flow is determined by a single intrinsic length scale. The role of this length scale thus reemerges from setting the decay length in the exponential eddy size distribution of the free turbulent regime, to determining the minimum size of circulating eddies featuring a scale-free power law. Our results demonstrate that soft-confinement provides with an invaluable tool to probe the intrinsic length and time scales of active materials, and pave the way for further exploration looking for similarities and differences between active and passive two-dimensional turbulence.

## Introduction

Many examples of either living or activated entities display intriguing modes of non-equilibrium behavior that cover a wide range of length scales and obey vastly different operational mechanisms. This general framework defines what is referred to as the realm of active matter (*1, 2*). One of the most distinctive collective features in this context is the emergence of "meso-scale", or active turbulence (*3-7*). This is a transversal concept that has been applied to characterize systems ranging from bacterial baths (*8, 9*) to *in vitro* aqueous reconstitutions of cytoskeletal proteins (*10-12*). Different from classical inertia-based turbulence in Newtonian fluids, active turbulence is dominated by dissipation. Both types of turbulence, however, share the disparity of spatial scales and directions of motions in the abundant jets and swirls that characterize turbulent flows. Recent numerical studies on the topology of turbulence in active nematics (AN), a two-dimensional realization of active gels featuring long-range orientational order, conclude that active turbulence is a multiscale phenomenon (*7*), characterized by an exponential distribution of vortices whose sizes scale in terms of an intrinsic, activity-dependent length, $l_\alpha \sim \sqrt{K/\alpha}$. This active length scale fixes the range over which active and elastic stresses are balanced (*7, 13-16*), and can be varied by tuning the activity ($\alpha$) or the rigidity ($K$) of the AN.

In spite of recent theoretical (*17-19*) and experimental (*20-22*) attempts, less attention has been dedicated to analyze the role of confining conditions on active fluids. This is an important question in a cell biology context, for instance in relation to cytoplasmic streaming in eukaryotes (*23*). Here, we show that active turbulence adapts to soft lateral confinement conditions that are specifically designed to preserve the integrity of the active material. Flow patterns display scale free characteristics, as the distribution of active vortex sizes changes from exponential into a power law at scales larger than a material-dependent length scale, which we identify with $l_\alpha$. This quantity thus reappears as an essential feature of the AN that serves to establish a minimum setting for active flows under soft confinement.

The scenario we experimentally study is that of a fully developed microtubule-based AN in contact with a soft interface patterned with self-assembled microdomains. We take advantage of the hydrodynamic coupling between the active and passive fluids across the water/oil interface and the tendency of some hydrophobic oils to form lamellar phases (Smectic-A liquid crystal phase) that can spontaneously organize in circular domains with giant viscous anisotropy. We observe that active flows are circularly trapped by large domains at the interface, while small domains can only act as scattering centers for the quasi-two-dimensional turbulent currents. By changing some control parameters of the active material, we demonstrate that the threshold length scale separating trapping from scattering is related to the bending rigidity and to the activity, as suggested by existing scaling predictions for the active length scale (*7, 15*).

# Results

## Active turbulent regime

The material we consider here is constituted by fluorescently labeled, micron-size, microtubules (MTs), bundled together by the depleting action of polyethylene glycol (PEG), and cross-linked by clusters of kinesin motor proteins (*10*). Under continuous supply of adenosine triphosphate (ATP), the tens-of-microns-long filamentous bundles are permanently subjected to internal stresses, which are triggered by the local shear forces that the motor complexes exert. This aqueous preparation spontaneously self-assembles into a flow-permeated active gel (*24-26*). Turbulent-like behavior is made more apparent when condensing the active gel at fluid interfaces, forming the so-called active nematic, where correlations in long-range filament orientation develop within a film that is hundreds of nanometers thick. Continuous bundle stretching and folding result in characteristic textures, observable using fluorescent probes or taking advantage of the reflectivity contrast of the active mesophase. Similar to what happens in passive nematic liquid crystals (LCs), the textures of these active films are punctured with defects, which, under high activity conditions, proliferate giving rise to a two-dimensional active turbulent steady state, sustained by balanced rates of creation and annihilation of defect pairs (*27-30*).

As a reference experiment, a layer of AN is interfaced with an isotropic oil, which is placed in a custom-built cell featuring a cylindrical well (see Materials and Methods and Fig. S2). Initially, the oil rests on the hydrophilic glass plate support and it is open to the air at the other side. The hydrophilic substrate is subsequently wetted by the aqueous active phase when injected underneath the oil. When in contact through biocompatible, surfactant-decorated interfaces, the AN displays disordered dynamics that arises from interfilament sliding and folding, while active flows propagate from the active into the passive phase (Fig. 1A; see also Video S1). Defects can be easily identified in the micrographs as regions devoid of MTs around which active filaments are organized either with a parabolic (+1/2 defects) or a hyperbolic (-1/2 defects) arrangement. This configuration constitutes the active turbulent regime, whose structure we have characterized by analyzing the statistical distribution of eddy sizes.

In order to locate and measure the area of each eddy, the local instantaneous velocity of the active flow is evaluated from a sequence of micrographs. Velocimetry data are used to obtain values of the Okubo-Weiss parameter, $OW = (\partial_x v_x)^2 + \partial_y v_x \cdot \partial_x v_y$, which provides with a standard criterion for eddy location, often used in fluid dynamics, by considering the extension of each eddy to be bound by the condition $OW < 0$ (*7, 31*) (Figs. 1B and 1C). By extracting this parameter from the experimental images, we are able to find the population of eddies during the observation of a given region of the active turbulent material for an extended period of time. A statistical analysis reveals an exponential distribution of eddy sizes (Fig. 1D), consistent with the existence of a characteristic length scale. This result agrees with the analysis reported in (*7*) using numerical simulations of the active turbulent regime in 2d nematodynamics, where this length scale is identified with $l_\alpha$.

## Conditioning the active flow with a patterned interface

The rheology of the water/oil interface is drastically changed when an anisotropic oil is used in contact with the AN. Here, we use the hydrophobic oil octyl-cyanobiphenyl (8CB), which was chosen because it features liquid crystal (LC) phases at temperatures compatible with protein activity (See Materials and Methods). The water/LC interface is stabilized with a PEG-based triblock copolymer surfactant, which both promotes the depletion of the active gel and the alignment of the 8CB molecules parallel to the interface, which will be crucial to achieve the desired surface patterning. Real time observation is performed using fluorescence, polarization and confocal microscopies (See Materials and Methods). A temperature quench below 33.4°C triggers the reversible transition of 8CB into the Smectic-A (SmA) phase, characterized by a microscopic organization in monomolecular-thick lamellae with molecules oriented along the local normal to the lamellar planes. Spontaneously, 8CB molecules align parallel to the polymer-decorated water/LC interface, and perpendicular to the LC/air boundary. In the absence of external fields, free energy minimization compatible with these hybrid anchoring conditions results in the SmA lamellae spontaneously self-assembling into polydisperse domains, known as *toroidal focal conic domains*, TFCDs (*32*). At the water/LC interface, TFCDs have a circular footprint and are formed by concentric SmA planes perpendicular to the interface. As a result, 8CB molecules, which are both parallel to the water layer due to the interaction with the surfactant, and perpendicular to the SmA planes, orient radially in concentric rings (Fig. 2A and Fig. S1). TFCDs organize into a fractal tiling known as *Apollonian gasket (32, 33)* that is able to cover the surface while satisfying the topological constraints imposed on the anisotropic oil. A characteristic of the SmA phase is that, although molecules diffuse freely within a given lamellar plane, flow is severely hindered in the direction perpendicular to the planes. As a consequence, the interfacial shear stress probed by the active material is markedly anisotropic. This anisotropy, in combination with the strong hydrodynamic coupling across the water/LC interface, forces active stretching of the MT bundles to occur preferentially along circular trajectories, centered in TFCDs. Swirling currents characteristic of the active turbulent flow evolve within the interfacial domain limits, segregated from the rest of the large-scale flows in the system (Fig. 2B). This new configuration becomes more apparent when looking at a time-averaged image of the AN. Individual eddies are clearly visible (Fig. 2C; see also video S2), with a dark central core surrounded by a bright corona, both in registry with the contacting TFCD. The hydrodynamic coupling between the AN and the patterned anisotropic oil interface has an additional dramatic effect on the active flow geometry: the size distribution of eddies becomes commensurate with the Apollonian tiling of TFCDs. Since the latter features a power-law distribution of domain sizes, the original exponential eddy size distribution of the turbulent AN is transformed into a scale-free power law distribution (Fig. 2D).

## Intrinsic length scale of the constrained Active Nematic

The effectiveness of the hydrodynamic confinement of AN flow in circular eddies depends on the size of each TFCD. Trapping by large domains is effective, while active flow over small TFCDs locally maintains the signature of active turbulence. We thus observe the emergence of an intrinsic length scale that stablishes the cut-off above which TFCDs are able to imprint a S=+1 eddy on the local AN (shaded region in Fig. 2D). Active flow trapping can be better visualized by tracing the trajectories of +1/2 defects retained within TFCDs for a given duration of time (Fig. 3). While small domains can, at most, scatter moving defects, large enough domains are able to locally trap their trajectories. As a result, defects are forced to circulate around the center of large TFCDs, following the easy flow circular directions defined by the underlying layered structure of the SmA phase (Fig. 3A). The distribution between trapped and unbound trajectories depends on the state of the AN, as can be seen by comparing Figs. 3A and 3B (see also Video S3), the latter featuring a concentration of ATP ten times lower, which results in a less effective trapping of defects. In order to quantify the active flow confinement in a given TFCD, we have computed, for each +1/2 defect that hovers within its limits, a winding number parameter, $Q = \Delta\alpha/2\pi$, where $\Delta\alpha$ is the accumulated angle of rotation around the TFCD center until the defect annihilates or escapes from the domain. A value Q > 1 implies that the +1/2 defect has performed more than one full rotation around the center of the TFCD and, therefore, we will consider that it has been trapped. Contrarily, $Q < 1$ corresponds to defects whose motion is perturbed by the anisotropic interface, but without ever describing a full circular trajectory. For the reference experimental conditions ([ATP] = 700 μM, [PEG] = 0.8 % w/w) the critical domain diameter is roughly 30 μm. Larger domains (Fig. 3C) are able to capture the active flow, while smaller ones (Fig. 3D) do not significantly alter the defect trajectories. We have analyzed the influence of the concentration of ATP, which will directly modify the activity parameter $\alpha \sim \ln[ATP]$, and the concentration of the depleting agent PEG, which is expected to modify the bending rigidity (thus *K*) of the active bundles (*34*). The latter should become thicker, and thus stiffer, for higher concentration of PEG. We observe that the minimum TFCD size for trapping increases when [ATP] is lowered, as well as when [PEG] is raised. This is evidenced in Figs. 3E and 3F, where under modified material conditions, the corresponding distribution of winding numbers shifts to smaller *Q* values when compared to the reference conditions for a domain of similar size (Fig. 3C).

## Defect topology in constrained Active Nematic eddies

In order to understand the origin of the length scale that defines the minimum TCFD size for the effective *capture* of the moving filaments, we need to focus on how the AN self-organizes to form a circulating eddy. We find that the combination of topological constraints, which dictate that a minimum of two +1/2 defects must be present in a rotating eddy, together with the existence of the characteristic active length scale $l_\alpha$, which stablishes the minimum bending radius of the active MT bundles, leads to the requirement that a TFCD must have a diameter above a critical value for an effective active flow trapping. As a characteristic feature of the active turbulence, defects are permanently renovated, even their number may change to a small extend, continuously assembling and disassembling the core structure of each rotating eddy. From a topological perspective, the confined rotating filaments organize a singularity of total charge S=+1, which corresponds to a full rotation around the domain center. However, the spontaneous folding of the extensile AN filaments can only create +1/2 defects (at the tip of the fold) or -1/2 defects (at the tail of the fold). This topological mismatch poses an additional constraint on the number of AN defects that evolve within a TFCD. At all times, a dynamic balance must be established such that the arithmetic sum of charges inside a single domain adds up to one, as dictated by topology (Fig. 4; see also Video S4). Clearly, the minimum number of semi-integer defects required to organize a rotating eddy is two S=+1/2 defects (Fig. 4A). Additional defects may be present, but with an equal number of positive and negative charges, so that the net balance is +1 (Figs. 4B and 4C). Due to the structure of the active filaments around distortions, +1/2 defects move much faster than their negative counterparts, which are simply advected by the active flow (*27, 35*). For this reason, active flow is determined, and usually characterized, by the motion of +1/2 defects. Indeed, in the soft confinement conditions we report here, we observe that +1/2 defects move most of the time circularly at constant speed, while the negative defects accumulate at the center of the eddies. This results in a heterogeneous distribution of the trapped active filament density. For large TFCDs, a central region, which appears dark in the long exposure time fluorescence micrographs, is mostly devoid of MTs. This area contains the cores of both quasi-static negative and dynamically rearranging positive defects. At the periphery, defect cores are only found occasionally, so this region is mostly occupied by the outer arms of the parabolic filament bundles that form +1/2 defects. Consequently, the MT density inside rotating eddies is maximized in this outer corona centered in a given TFCD, as observed in Fig. 2C.

### Periodic bending instability of the aligned active nematic

The extensile nature of the studied AN material makes a configuration of parallel elongated filaments intrinsically unstable, leading to the defect-forming bending instability (*13, 14*). As a consequence, the inner structure of S=+1 eddies organized by TFCDs is periodically disrupted and reconstituted. An example of this process is depicted in Fig. 5 (see also Video S5), where we show an eddy, which has been assembled by a TFCD of approximate diameter 200 µm. In the first micrograph (Fig. 5A), the AN is structured as an annular band of circularly-aligned MT bundles. This configuration is unstable (Fig. 5B), and the filaments develop a radial buckling instability that leads to the formation of semi-integer defect pairs, moving inwards and outwards of the domain (Figs. 5C-5D). Some of these new defects will eventually annihilate with existing inner ones (Figs. 5E-5F), finally reorganizing the original aligned state (Fig. 5G). This sequence of dynamic events, which is also exhibited as a time periodic modulation of the velocity of the active flows inside domains, repeats with remarkable periodicity, thus revealing an intrinsic time scale of the active nematic [See Fig. S3 and Refs. (*15, 16*)].

# Discussion

There has been recently a surge of interest in active materials both of biological and non-biological nature. In many experimental realizations these fluids adopt marked trends of liquid crystalline order (*i.e.*, long-range orientational order), locally punctuated by defects that drive continuously permeating hydrodynamic currents. In particular, the system here considered, composed of bundled microtubules entangled by kinesin motors, *in vitro* reconstitutes the textures and flows that are present at a cellular level. We have shown here how a particularly well-recognized hallmark of active materials, the so-called active turbulence, adapts to confinement. We find that soft-*parceling* the AN with a tiled patterned interface prepared with the anisotropic oil 8CB has a large influence on the distribution of the active velocity and vorticity fields.

It is well accepted in the literature that the spatial arrangement of ANs is determined by the single intrinsic length scale $l_\alpha$, which establishes spatial features in the material such as the steady-state defect separation and the bending radius of the extensile MT bundles. Earlier simulations revealed that this length scale also determines the geometry of the active turbulent regime (*7*), where the eddy size distribution follows an exponential distribution whose decay length is related to $l_\alpha$. Our experiments confirm these findings when the AN evolves in contact with an isotropic oil (Fig. 1). On the other hand, when flowing in contact with the high anisotropy, patterned SmA phase of 8CB, the active turbulence is reorganized in a scale-free power law distribution of eddy sizes that mimics the underlying SmA tiling. However, we have put into evidence that, even in this regime, an intrinsic length scale resurfaces in the form of a minimum eddy size compatible with AN flow organization. We have provided topological arguments that justify the existence of this minimum size, since a minimum of two +1/2 defects (parabolic folds) of the active filaments are required to organize a circulating eddy, and the defect separation is regulated by $l_\alpha$ (*7, 15, 27-29*). We have explored the dependence of the minimum eddy size on [ATP] (related to the activity, $\alpha$) and on [PEG] (related to the bending rigidity of the active bundles, $K$). Although the range of usable values for [ATP] is rather limited for this system, and $K$ will increase with [PEG] but with an unknown functional dependence, the observed scaling of this characteristic size is consistent with the one expected for $l_\alpha$. This evidence, along with the topological arguments presented above, let us summarize our findings as a manifestation of the different roles played by the active length scale $l_\alpha$, depending on the confinement constraints. For isotropic or quasi-isotropic surface rheology, the *free* active turbulence is characterized by an exponential distribution of eddy sizes, with $l_\alpha$ being related to the decay length. Contrarily, in the case of the AN in contact with the circular SmA domains, *constrained* active flow self-organizes by matching the scale-free distribution of underlying TFCDs. In this case, $l_\alpha$ appears as the size threshold separating entrapping from scattering domains.

In addition, the ability to entrap the flows of active materials provides with a useful tool to constrain their dynamics and study new organizational features arising in bounded conditions. In this work, by confining the AN in circulating eddies, we have been able to explore the dynamical stability of these arrangements, which, because of the extensile nature of their constituents, are metastable. In spite of the flow reorganization triggered by episodes of structural instability, the handedness of flow rotation inside each eddy, which is randomly selected upon self-assembly, is preserved in time. The flows generated by these rotating eddies are able to drag passive tracer beads, which circulate together with the active material. As they are not bound to a circular trajectory, the tracers inside rotating eddies are dragged outwards while performing spiral trajectories. On the other hand, particles adsorbed at the LC phase describe a perfect circular trajectory, following the geometry of the SmA planes (see Fig. S3 and Video S6).

Finally, our findings should stimulate further theoretical and experimental studies dedicated to investigate the similarities and differences between classical and active two-dimensional turbulence. Significantly, in this protein-based preparation, energy is injected at the lowest possible length scale, represented by the molecular hydrolysis of ATP. This chemical energy is transferred to larger length scales through two different, and intimately linked, mechanisms: the standard kinetic energy route involving dissipation, and via elastic stresses storage in the continuously reforming extensile material. The peculiarities that this remarkable new feature of active soft matter might be unveiled when addressing the energy vs. enstrophy dual cascade, commonly investigated in two-dimensional inertial turbulence (*36*), promise striking future discoveries on the real nature of active flows.

# Materials and Methods

## Materials

**Lamellar liquid crystal (LC)**
4-cyano-4'-octylbiphenil (8CB, Synthon; ST01422) is a thermotropic liquid crystal between 21.4 and 40.4ºC, featuring a lamellar Smectic-A phase in the temperature range 21.4 ºC < T < 33.4 ºC.

**Active gel**
Polymerization of microtubules (MTs). Heterodimeric (α,β)-tubulin (from bovine brain, obtained from the Brandeis University Biological Materials Facility) is incubated at 37ºC for 30min in an M2B buffer (80mM PIPES, 1mM EGTA, 2mM $MgCl_2$) (Sigma; P1851, E3889 and M4880, respectively) supplemented with the reducing agent dithiothrethiol (DTT) (Sigma; 43815) and with Guanosine-5'-[(α,β)-methyleno]triphosphate (GMPCPP) (Jena Biosciences; NU-405), a non-hydrolysable analogue of the nucleotide GTP. As GTP, GMPCPP promotes the association between tubulin heterodimers although it completely suppresses the dynamic instability of MT (*37*). By controlling the concentration of GMPCPP we are able to prepare high-density suspensions of short MTs (1-2 µm). For the characterization with fluorescence microscopy, part of the initial tubulin (3%) has to be fluorescently labelled.

Kinesin expression. Drosophila Melanogaster heavy chain kinesin-1 K401-BCCP-6His (truncated at residue 401, fused to biotin carboxyl carrier protein (BCCP) and labelled with six histidine tags) has been expressed in Escherichia coli by using the plasmid WC2 from The Gelles Laboratory (Brandeis University) and purified with a nickel column (*37*). After dialysis against 500mM Imidazole buffer, kinesin concentration is estimated by means of absorption spectroscopy and stored at a specific concentration in a 40% wt/vol sucrose solution at -80ºC (*37*).

Preparation of molecular motor clusters. Biotinylated kinesin motor proteins and tetrameric streptavidin (Invitrogen; 43-4301) are incubated on ice for 30 minutes at specific stoichiometric ratio (~2:1) in order to obtain kinesin-streptavidin motor clusters.

Assembly of the active gel. MTs are mixed with the motor clusters that act as cross-linkers, and with ATP (Sigma; A2383) that drive the activity of the gel. The aqueous dispersion contains a non-adsorbing polymer (poly-ethylene glycol, PEG, 20 kDa) (Sigma; 95172) that promotes the formation of filament bundles through depletion interaction (Fig. S2). To maintain a constant concentration of ATP during the experiments, an enzymatic ATP-regenerator system is used, consisting on phosphoenolpyruvate (PEP) (Sigma; P7127) that fuels pyruvate kinase/lactate dehydrogenase (PK/LDH) (Invitrogen; 434301) to convert ADP back into ATP. Several antioxidant components are also included in the solution to avoid protein denaturation, and to minimize photobleaching during characterization by means of fluorescence microscopy. The PEG-based triblock copolymer surfactant Pluronic F-127 (Sigma; P-2443) is added at 1% (wt/wt) (final concentration) to procure a biocompatible water/oil interface in subsequent steps.

## Methods

**Experimental Setup**
The studied AN is formed at the interface with either an isotropic silicone oil (BlueStar Silicones; BlueSil® V47:12500) or 8CB. The interface is prepared in a cylindrical well of diameter 5 mm and depth 2 mm, manufactured with a block of poly-dimethylsiloxane (PDMS) (Sylgard®, Dow Corning) using a custom mold. The block is glued onto a bioinert and superhydrophilic polyacrylamide (PAA)-coated glass, which is prepared following (*38*) (see Fig. S2). In brief, clean and activated glass is first silanized with an acidified ethanolic solution of 3-(Trimethoxysilyl)propylmethacrylate (Sigma; 440159), which will act as polymerization seed. The silanized substrates are subsequently immersed in a aqueous solution of acrylamide monomers (Sigma; 01697) (2 %wt/vol, for at least 2 h) in the presence of the initiator ammonium persulphate (Sigma; A3678) and N,N,N',N'-Tetramethylethylenediamine (Sigma; T7024), which catalyzes both initiation and polymerization of acrylamide. The cavity is first filled with 50 µL of oil and, subsequently, 1 µL of the water-based active gel is injected between the hydrophobic oil and the superhydrophilic glass plate (Fig. S2). Pluronic F-127 stabilizes the interface and avoids direct contact between the oil and the protein-based active mixture. Moreover, this surfactant ensures a planar alignment of 8CB molecules at the water/LC interface. When 8CB is used, samples are heated up to 35 °C to promote transition to the less viscous nematic phase of the LC, which facilitates the spreading of the active gel onto the PAA-coated substrate. Temperature is controlled by using a thermostatic oven built with Thorlabs SM1 tube components and tape heater, which is regulated with a Thorlabs TC200 controller. After several minutes at room temperature, the active material in the gel spontaneously condenses onto the flat water/oil interface, leading to the formation of the AN layer. Unlike conventional flow cells, in which a layer of the active gel is confined in a thin gap between two glass plates, our open setup enables us to prepare the interface using oils with a high viscosity.

**Imaging**
Routine observations of the AN were performed by means of conventional epifluorescence microscopy. We used a custom-built inverted microscope with a white led light source (Thorlabs MWWHLP1) and a Cy5 filter set (Edmund Optics). Image acquisition was performed with a QImaging ExiBlue cooled CCD camera operated with ImageJ μ-Manager open-source software. For sharper imaging of the interfacial region, we used a Leica TCS SP2 laser-scanning confocal microscope equipped with a photomultiplier as detector and a HeNe–633-nm laser as light source. A 20x oil immersion objective was employed. We performed confocal acquisition in fluorescence and reflection modes, simultaneously.

**Image analysis**
Tracer-free velocimetry analysis of the AN was performed with a public domain particle image velocimetry (PIV) program implemented as an ImageJ plugin (*39*). Flows were also traced by dispersing PEGylated spherical polystyrene microbeads of diameter 1.7 μm (Micromod; 08-56-173). Manual Tracking ImageJ plugin was used to manually track trajectories of particles or defects in the AN. Further analysis of velocimetry data was performed with custom-written MatLab codes. In order to quantify the vortex distribution in the turbulent AN, the Okubo-Weiss (OW) parameter was mapped from the velocity fields obtained by image velocimetry (Fig. 1). The vortex areas were quantified from thresholded OW field binary images in ImageJ. Vortices on the edges of the field of view were excluded.

# Supplementary Materials

Fig. S1. Structure of the TFCDs.
Fig. S2. Sketch of the experimental setup.
Fig. S3. Flows induced by active nematic eddies.
Video S1. Turbulent active nematic.
Video S2. Active nematic eddies under toroidal focal conic domains (TFCDs).
Video S3. Influence of the activity on the defect trajectories of confined active nematics.
Video S4. Preservation of the topological charge inside active vortices.
Video S5. Active nematic eddies are metastable.
Video S6. Transition from constrained to unconstrained active turbulence.

# Acknowledgments

**General**: The authors are indebted to Z. Dogic and S. DeCamp (Brandeis University), and the Brandeis University MRSEC Biosynthesis facility for their assistance in the preparation of the active gel. We thank B. Hishamunda (Brandeis University), and M. Pons, A. LeRoux, and G. Iruela (Universitat de Barcelona) for their assistance in the expression of motor proteins. We thank R. Casas and G. Valiente (Bluestar Silicones) for providing the silicone oil sample.

**Funding:** Funding has been provided by MINECO (project FIS 2013-41144P). P.G. acknowledges funding from Generalitat de Catalunya through a FI-DGR PhD Fellowship. Brandeis University MRSEC Biosynthesis facility is supported by NSF MRSEC DMR-1420382.

**Author contributions:** All authors conceived the research. P. G. performed the experiments and data analysis. J. I.-M. and F. S. supervised the research and wrote the manuscript. All authors discussed the results.

**Competing interests:** The authors declare no competing financial interests.

# Figures

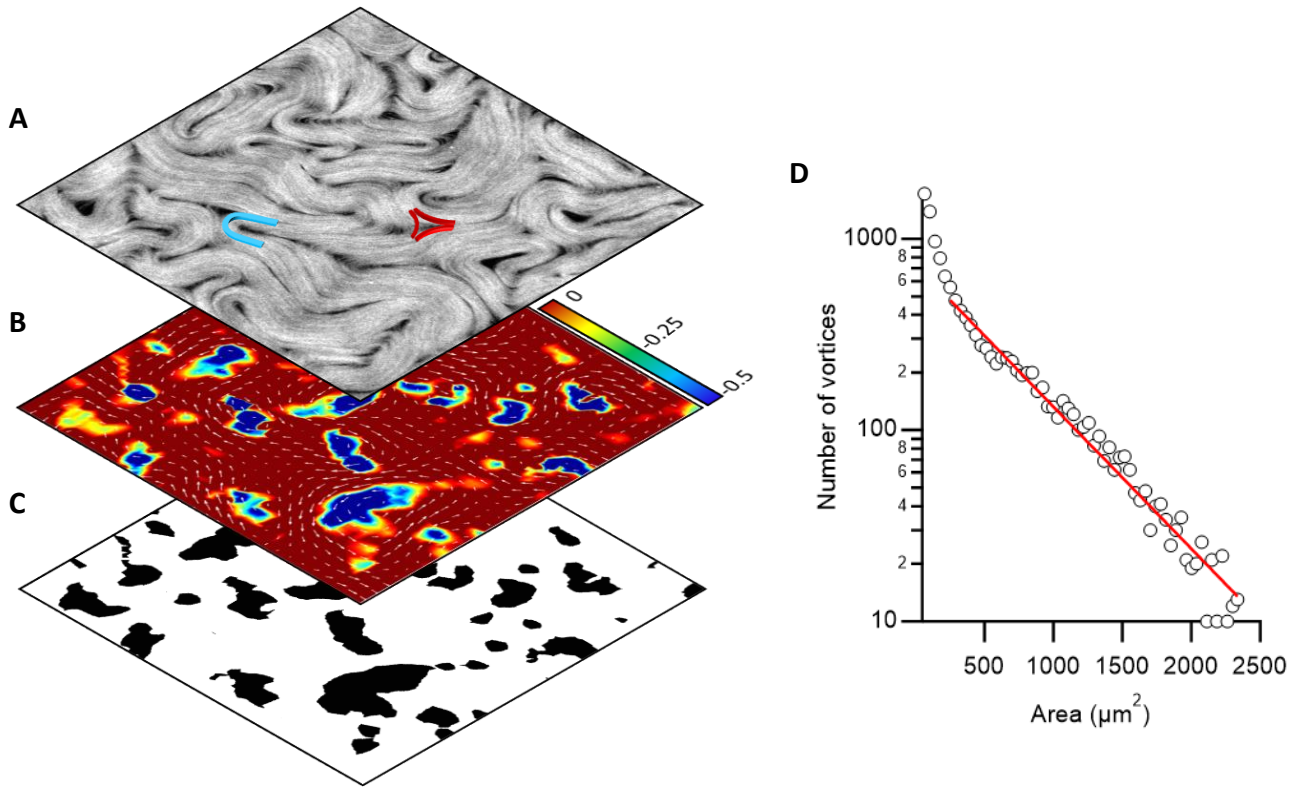

**Figure 1. Structure of the active turbulent flow.** (A) Confocal fluorescence micrograph (375×375 µm$^2$) of the active nematic in the turbulent regime when in contact with an isotropic oil. An example of the proliferating +1/2 (blue) and -1/2 (red) defects is sketched on top of the image. (B) Instantaneous flow field (vector plot) and computed Okubo-Weiss parameter field (density plot, arbitrary units), which is employed to determine the location and size of eddies. (C) Binary image corresponding to the Okubo-Weiss field. (D) Statistical analysis of the distribution of eddy sizes accumulated for a series of one thousand snapshots during the active turbulent flow. The solid line is an exponential fit to the data. The range of sizes is limited by the size of the field of view. (See Video S1).

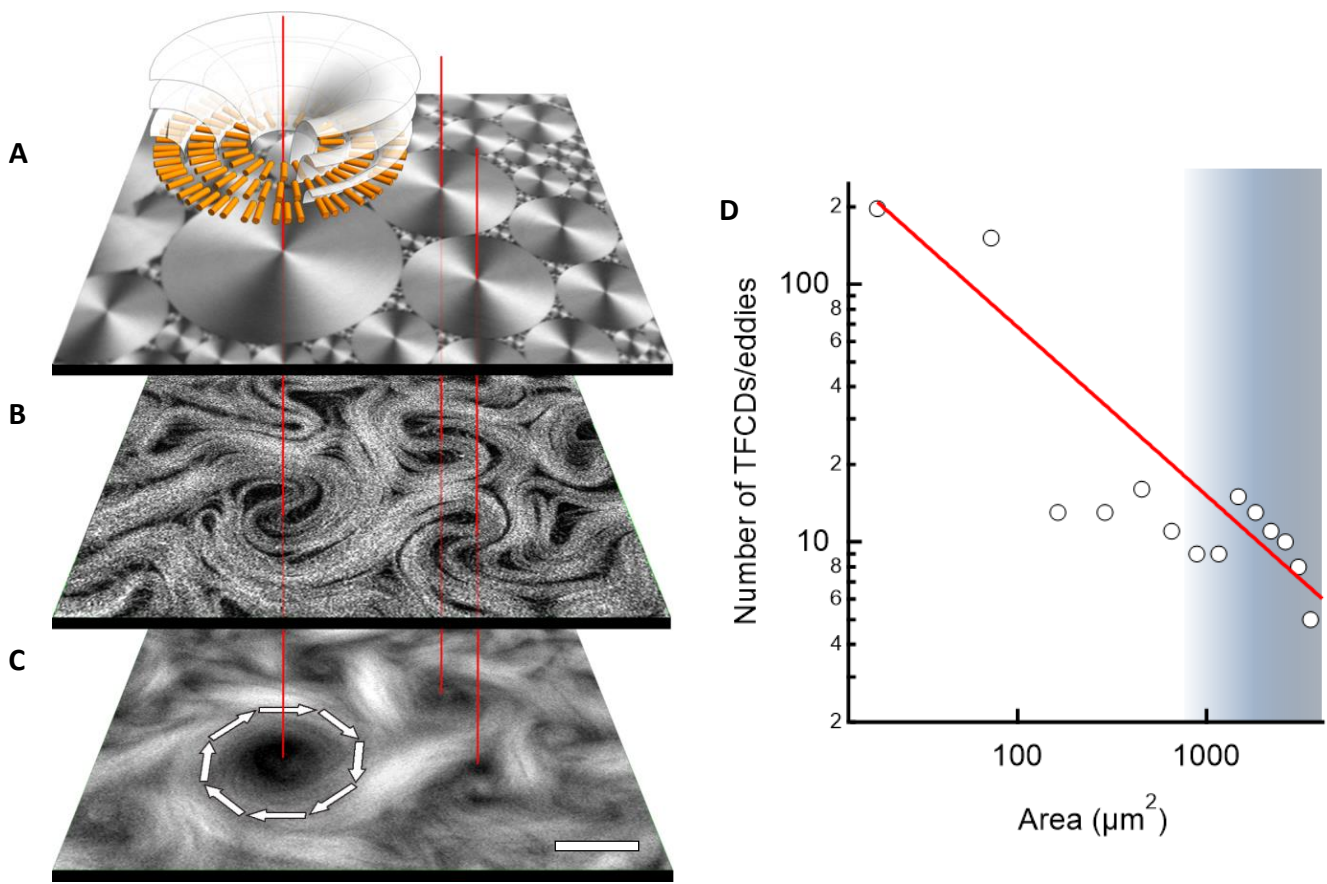

**Figure 2. Self-assembly of the active nematic in contact with the patterned interface.** (A) Confocal reflection micrographs of the oil/water interface reveal the tiling formed by the distribution of circular TFCDs of the liquid crystal 8CB in the Smectic A phase. The diagram illustrates the arrangement of the mesogen molecules at the interface. (B) Confocal fluorescence micrograph showing the rearrangement of the active nematic flow due to the hydrodynamic coupling with the anisotropic interface. (C) Time averaged fluorescence confocal micrograph (total integration time 300 s) highlights the patterning of the active flow due to the anisotropic interface. Arrows indicate the direction of the circular flow. Line segments through the panels identify coincident regions. Scale bar, 25 µm. (See Video S2). (D) Analysis of the size distribution of TFCDs in a 500×500 µm$^2$ window. The line is a power law fit to the data with an exponent -0.7. The shaded region corresponds to the range of domains that are able to effectively trap the active flow. The threshold can be varied with the experimental control parameters (see text and Fig. 3).

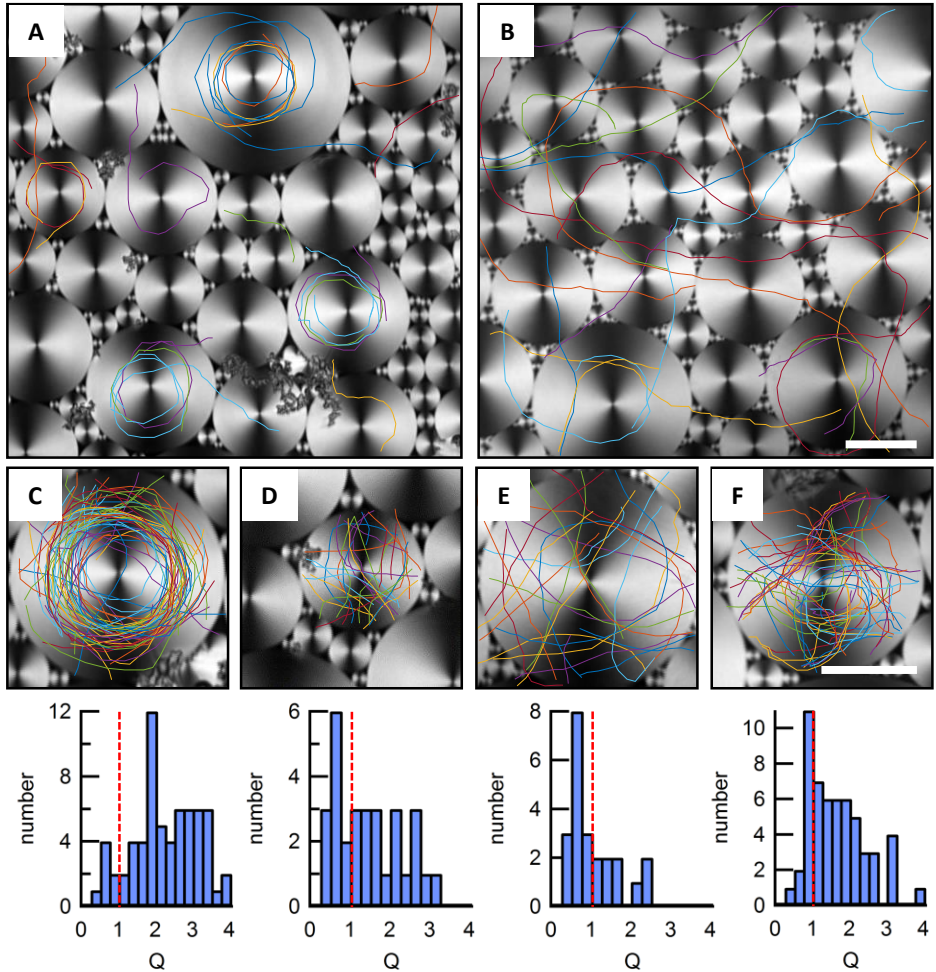

**Figure 3. Soft confinement of the active nematic.** Confocal reflection micrographs of different TFCDs and overlaid traces of +1/2 defects of the flowing active nematic for different experimental conditions. Histograms below micrographs (C-F) include the distribution of computed winding numbers, $Q$ (see text), corresponding to defect trajectories shown in the respective micrographs. The dashed line indicates $Q = 1$, which implies a full rotation. Experimental conditions are: (A, C, and D) [ATP] = 700 µM, [PEG] = 0.8 %w/w. (B and E) [ATP] = 70 µM, [PEG] = 0.8 %w/w. (F) [ATP] = 700 µM, [PEG] = 4.2 %w/w. Scale bars, 25 µm. (See Video S3).

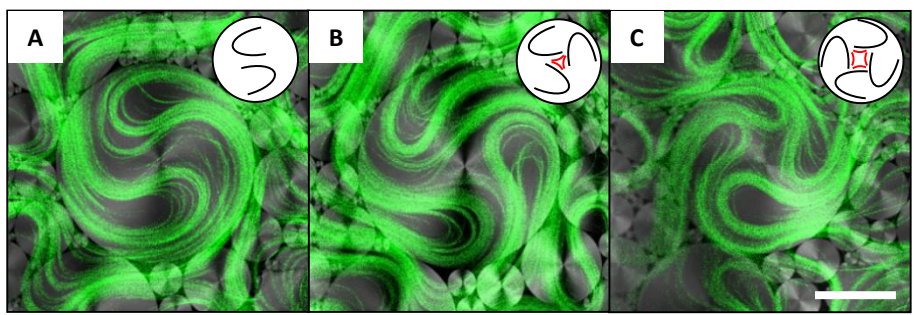

**Figure 4**. **Defect structure in active nematic eddies.** Different confocal micrographs of the active nematic in contact with the lattice of TFCDs. The confocal fluorescence (green) and confocal reflection (grayscale) channels are overlaid and show, respectively, the active nematic filaments and the SmA interface. The active flow trapped by the central domain features different topological defect configurations, as illustrated by the sketches on the top right corners, always with a net topological charge balance S = +1. (A) S = 2×(+1/2) ; (B) S = 3×(+1/2)-1/2; (C) S = 4×(+1/2)-1. Scale bar, 50 µm. (See Video S4).

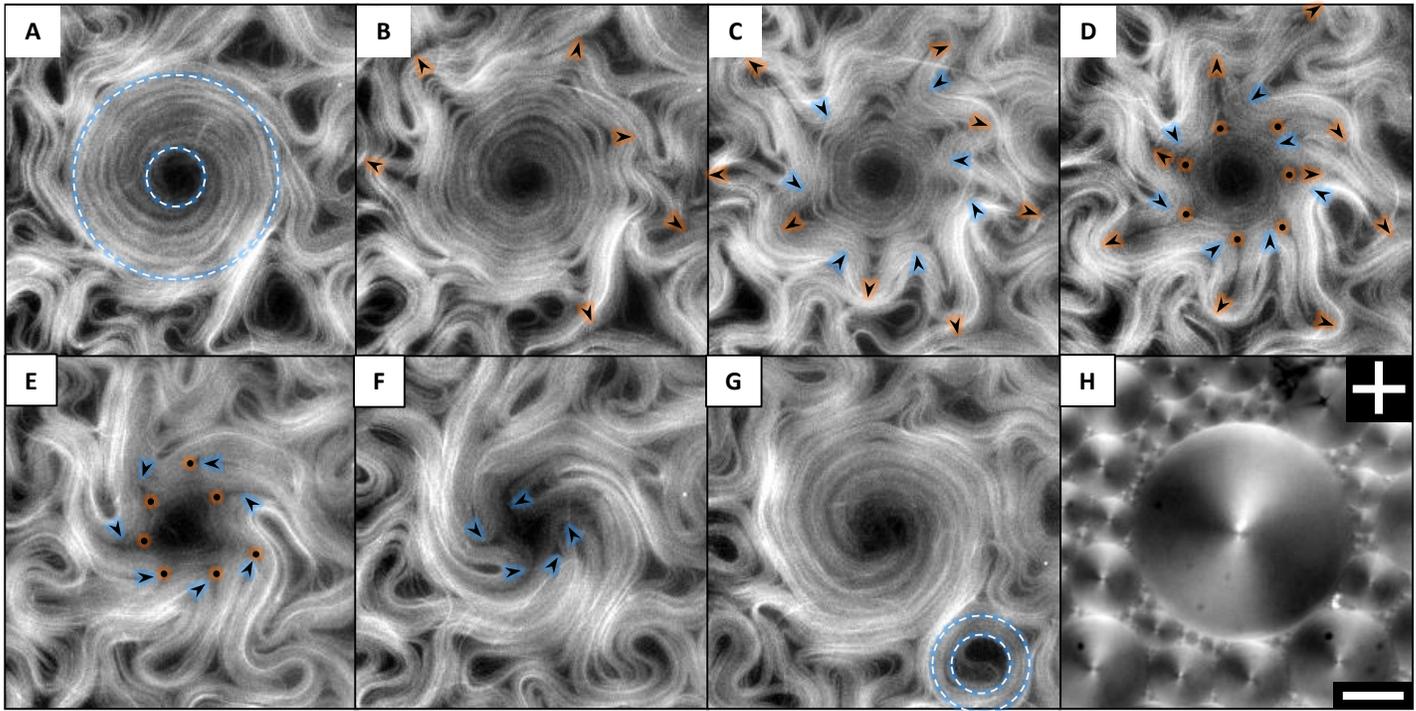

**Figure 5. Defect dynamics in active nematic eddies.** Fluorescence confocal micrographs show the periodic reconstruction of the active nematic eddy constrained by a TFCD. (A) Rotation of +1/2 defects results in the formation of a corona of MT bundles, which align in quasi-concentric circumferences. Dashed lines depict inner and outer corona perimeters. (B-D) radial bending instability of the aligned bundles generates pairs of complementary +1/2 (black arrow heads, blue incoming, red outgoing) and -1/2 (black dots highlighted in red) defects. (E-G) Incoming +1/2 defects annihilate with static -1/2 defects, leading to the formation of a new eddy structure. In (G), a smaller TFCD has assembled a MT corona. For clarity, -1/2 defects are only highlighted in D and E. (H) Polarizing micrograph taken between crossed polarizers (top right corner) of the local distribution of TFCDs that leads to the active flow patterns in (A-G). Scale bar, 50 μm. (See Video S5).

# Taming active turbulence with patterned soft interfaces

P.Guillamat, J.Ignés-Mullol, F.Sagués

**Table of contents**

1. **Supplementary Figures**

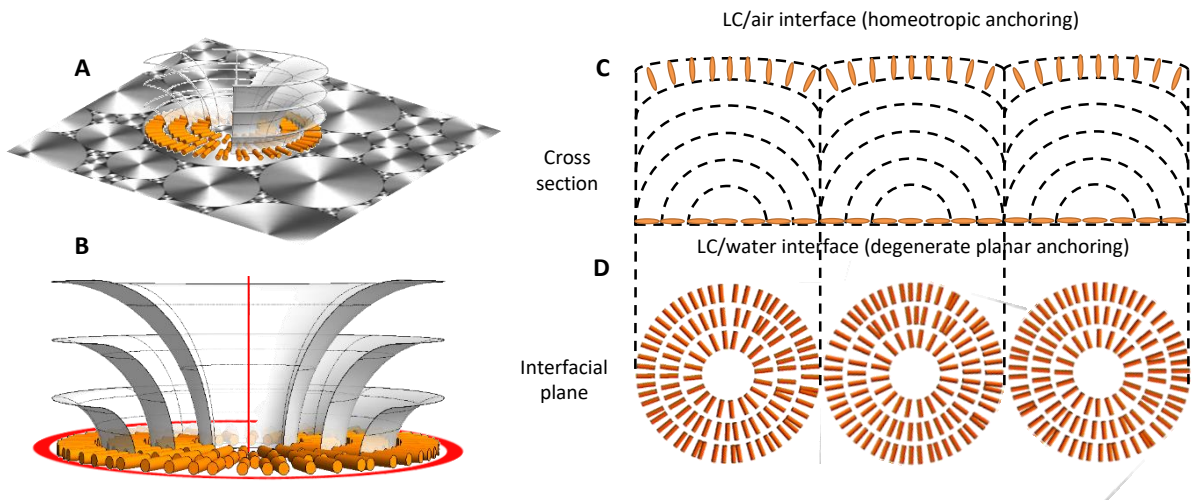

**Figure S1: Structure of the TFCDs.** (A) In our experiments, 8CB is subjected to hybrid boundary conditions as it is interfaced by air and water, which induce homeotropic (perpendicular) and (degenerate) planar anchoring of the LC, respectively. Under these conditions, 8CB arranges in TFCDs when it features the SmA phase. (B) SmA lamellae bend around two singular lines (depicted in red in the sketch) namely, a circle at the interface where the LC features planar anchoring, and a straight line running from the circle center into the bulk of the LC, where 8CB presents homeotropic alignment. (C) Cross section of the bulk structure of the lamellar planes in the LC across the sample thickness. (D) At the LC/water interface, TFCDs organize concentric SmA planes, where 8CB molecules orient radially, both parallel to the oil/water interface and perpendicular to the SmA planes.

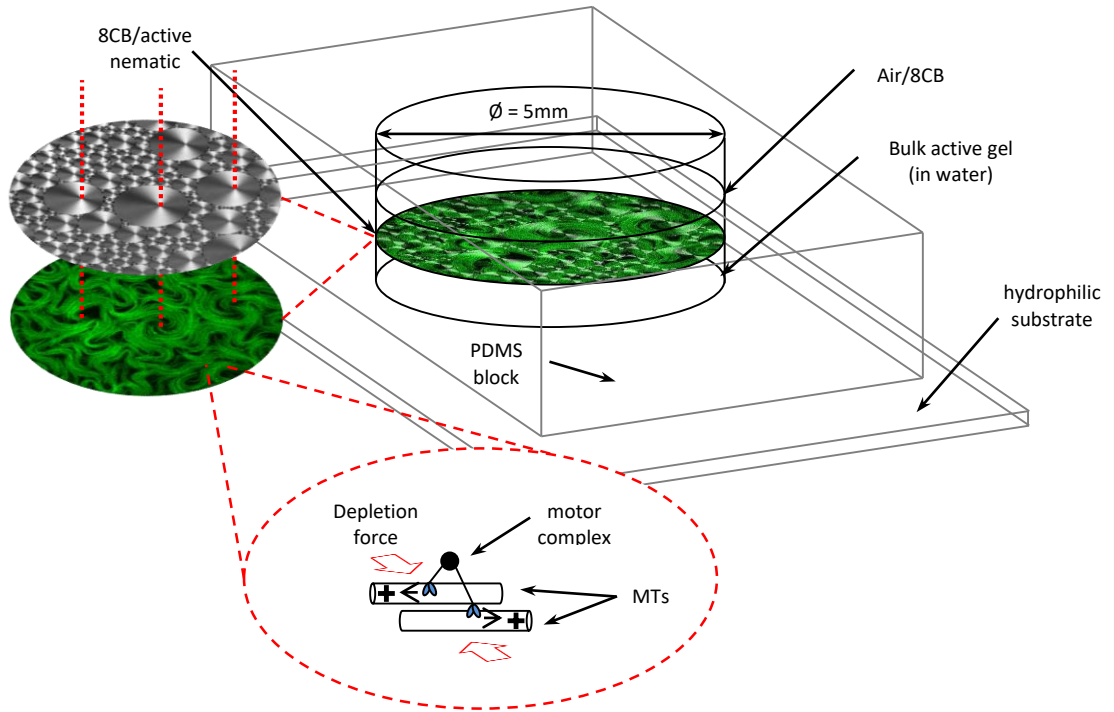

**Figure S2: Sketch of the experimental setup.** A block of polydimethylsiloxane (PDMS) with a cylindrical cavity (diameter = 5 mm) is glued onto a superhydrophilic glass plate, coated by a polyacrylamide brush. The cavity is first filled with 8CB and subsequently, an active aqueous gel is added underneath, where it spreads covering practically the whole base of the cylinder. The active gel is composed of microtubule bundles that are sheared by ATP-activated molecular motor clusters. Via depletion forces, the active material is driven through water towards the interface with 8CB, where it organizes an active nematic. In contact with the anisotropic interface, the active material forms eddies in order to adapt to the rheological constraints imposed by the interface.

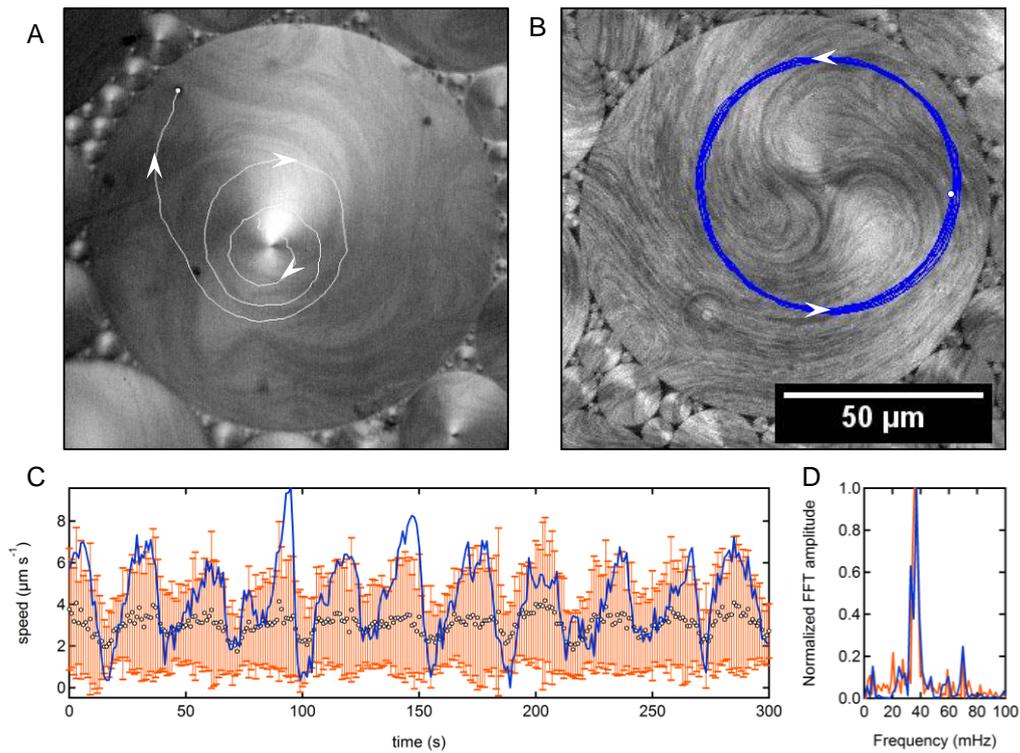

**Figure S3: Flows induced by active nematic eddies.** In contact with big enough TFCDs, the active nematic generates swirling vortices, which can effectively drag dispersed passive tracers (A, white dot) in spiral trajectories (white line). (B) Particles adsorbed at the oil phase move in perfect circles within the lamellar planes (blue line). Arrows indicate the sense of the rotation. (C) Active nematic eddies are unstable and they assemble and disassemble with striking regularity. Here, we plot the average speed of the active flows in (B) (white dots with orange error bars) together with the speed of the rotating tracer at the lamellar phase (blue line). (D) Normalized power spectra for data in (C) (See Video S6).

2. **Supplementary Videos**

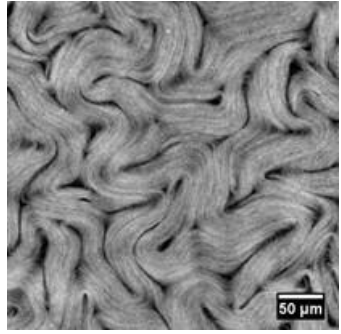

**Video S1: Turbulent active nematic.** Fluorescence micrographs of an active nematic in contact with an isotropic oil.

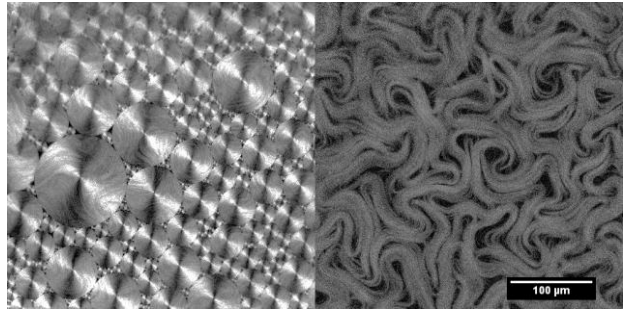

**Video S2: Active nematic eddies under toroidal focal conic domains (TFCDs).** Reflection confocal (left) and fluorescence (right) micrographs show how, in contact with big enough TFCDs, the active nematic organizes eddies, led by the circular patterns at the oil/water interface. Notice that small focal conics do not trigger the formation of eddies.

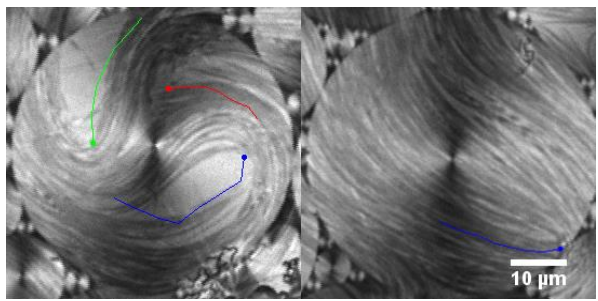

**Video S3: Influence of the activity on the defect trajectories of confined active nematics.** Reflection confocal micrographs with superimposed +1/2 defect trajectories evidence the difference between constrained active nematics in different activity conditions. At the interface with domains of equal size, the active nematic with high activity (left) gives rise to more circulation of defects. Defects in the low-activity active nematic (right) are only scattered by the circular patterns at the interface.

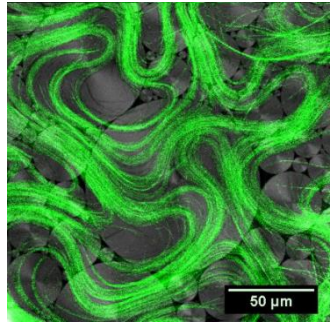

**Video S4: Preservation of the topological charge inside active vortices.** The video contains superimposed images of fluorescence and reflection confocal microscopy centered on a large active nematic eddy. The active nematic (in green) is forced to follow the concentric direction of the TFCDs in the oil phase (grayscale), with an accumulated defect charge constrained by the topology of a full circle (S=+1). The total number of defects varies but this topological constraint is satisfied at all times.

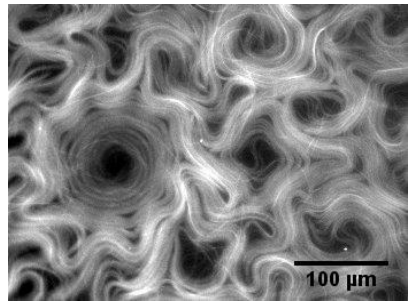

**Video S5: Active nematic eddies are metastable.** Fluorescence micrographs show how rotation of +1/2 defects organize a corona of MT bundles at the periphery of big eddies. Due to the extensile nature of the active material this ordered state is prone to suffer the bending instability, which disassembles the eddies by creating defects in the radial direction. New defects adapt to the rheological constraints at the interface and start to rotate again to reconstitute a new corona of MT bundles.

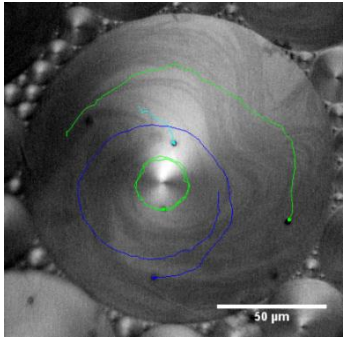

**Video S6: Active nematic eddies induce localized vortical flows.** Bright field micrographs of TFCDs in contact with the active nematic with polystyrene spherical particles (1.7 µm) suspended inside. The superimposed trajectories evidence how the flows induced by the active nematic eddies effectively drag the particles towards the periphery of the interfacial domains in spiral trajectories. Notice that one of the tracers (near the center of the TFCD and whose trajectory has been highlighted in green) moves in perfect circles. In this case, the tracer is moving adsorbed at the lamellar planes, which flow within the TFCD.